\documentclass[aps,reprint,superscriptaddress,notitlepage,longbibliography]{revtex4-1}
\usepackage{graphicx}
\usepackage{amsmath}
\usepackage{amsfonts}
\usepackage{bm}
\usepackage{color}
\usepackage{units}
\usepackage{ulem}

\begin{document}

\title{Experimental Realization of a Minimal Microscopic Heat Engine}

\author{Aykut Argun}
\thanks{These two authors contributed equally}
\affiliation{Department of Physics, University of Gothenburg, SE-41296 Gothenburg, Sweden, EU}

\author{Jalpa Soni}
\thanks{These two authors contributed equally}
\affiliation{Department of Physics, University of Gothenburg, SE-41296 Gothenburg, Sweden, EU}

\author{Lennart Dabelow}
\affiliation{Theoretisch-Physikalisches Institut, Friedrich Schiller University Jena, Max-Wien-Platz 1, 07743 Jena, Germany, EU}

\author{Stefano Bo}
 \affiliation{Nordita, Royal Institute of Technology and Stockholm University, SE-10691 Stockholm, Sweden, EU}

\author{Giuseppe Pesce}
\affiliation{Department of Physics E. Pancini, University of Naples Federico II, via Cintia, 80126-I, Naples, Italy, EU}

\author{Ralf Eichhorn}
\affiliation{Nordita, Royal Institute of Technology and Stockholm University, SE-10691 Stockholm, Sweden, EU}

\author{Giovanni Volpe}
\affiliation{Department of Physics, University of Gothenburg, SE-41296 Gothenburg, Sweden, EU}

\begin{abstract}
Microscopic heat engines are microscale systems that convert energy flows between heat
reservoirs into work or systematic motion. 
We have experimentally realized a minimal microscopic heat engine.
It consists of a colloidal Brownian particle optically trapped in an elliptical potential well and simultaneously coupled to two heat baths at different temperatures acting along perpendicular directions. 
For a generic arrangement of the principal directions of the baths and the potential, the symmetry of the system is broken, such that the heat flow drives a systematic gyrating motion of the particle around the potential minimum. 
Using the experimentally measured trajectories, we quantify the gyrating motion of the particle, the resulting torque that it exerts on the potential, and the associated heat flow between the heat baths. 
We find excellent agreement between the experimental results and the theoretical predictions.
\end{abstract}	

\date{\today}

\maketitle

During the last two decades, the rapid development of stochastic thermodynamics has provided scientists with a framework to explore the properties of nonequilibrium phenomena in microscopic systems where fluctuations play a prominent role \cite{sekimoto2010stochastic,seifert2008stochastic,chetrite2008fluctuation,ritort2008nonequilibrium,jarzynski2011equalities,seifert2012stochastic,van2013stochastic,van2015ensemble}. 
The advancement of experimental techniques (in particular optical trapping and digital video microscopy \cite{jones2015optical}) has made it possible to experimentally study thermodynamics at the single-trajectory level \cite{wang2002experimental,liphardt2002equilibrium,collin2005verification,berut2012experimental}. 
These tools have been applied, e.g., to investigate the performances of molecular machines \cite{astumian2010thermodynamics,seifert2012stochastic}. 
Furthermore, microscopic heat engines (i.e. artificial microscopic systems that extract heat from the surrounding thermal bath(s) and turn it into useful work or systematic motion) have been proposed theoretically\cite{hondou2000unattainability,schmiedl2007efficiency,Filliger2007Brownian,bo2013entropic,stark2014classical,verley2014unlikely,fogedby2017minimal,bo2017driven} and realized experimentally \cite{blickle2012realization,quinto2014microscopic,martinez2016brownian,krishnamurthy2016micrometre,schmidt2017microscopic,ciliberto2017experiments,martinez2017colloidal}, providing insights on fundamental aspects of non-equilibrium thermodynamics.

In this Article, we experimentally realize and investigate a minimal microscopic engine constituted of a Brownian particle held by a generic potential well and simultaneously coupled to two heat baths at different temperatures acting along perpendicular directions so that a non-equilibrium steady state is maintained. For a generic arrangement of the principal directions of the baths and the potential, the symmetry of the system is broken, such that the heat flow between the two heat baths drives a systematic gyrating motion of the particle around the potential minimum.
Originally, this engine was proposed theoretically by Filliger and Reimann \cite{Filliger2007Brownian};
it is considered to be minimal because of its intrinsic simplicity,
yet generating a torque via circular motion,
and because it works autonomously in permanent simultaneous contact with two heat baths (i.e., without the need for an external driving protocol).

In the experiment, we use a single colloidal particle suspended in aqueous solution at room temperature and trap it in an elliptical optical potential. 
The \textit{per se} isotropic thermal environment is rendered anisotropic by applying fluctuating electric signals with an almost white frequency spectrum along a specific direction; such techniques \cite{martinez2013effective,mestres2014realization,dinis2016thermodynamics} and similar ones \cite{gomez2010steady,berut2016stationary} have recently been demonstrated to generate in excellent approximation high temperature thermal noise with negligible friction effects.
In addition to experimentally confirming the prediction of Ref.~\cite{Filliger2007Brownian} for
the torque (see eq.~\eqref{rotat}), we characterize the gyrating motion of the colloid in more detail
by measuring the cross-correlation between the spatial coordinates.
Moreover, we analyze the energy exchanges between the two heat baths mediated by the particle's motion,
using the tools of stochastic energetics \cite{sekimoto1998langevin,sekimoto2010stochastic}.

\begin{figure*} [ht!]
\includegraphics[width=\textwidth]{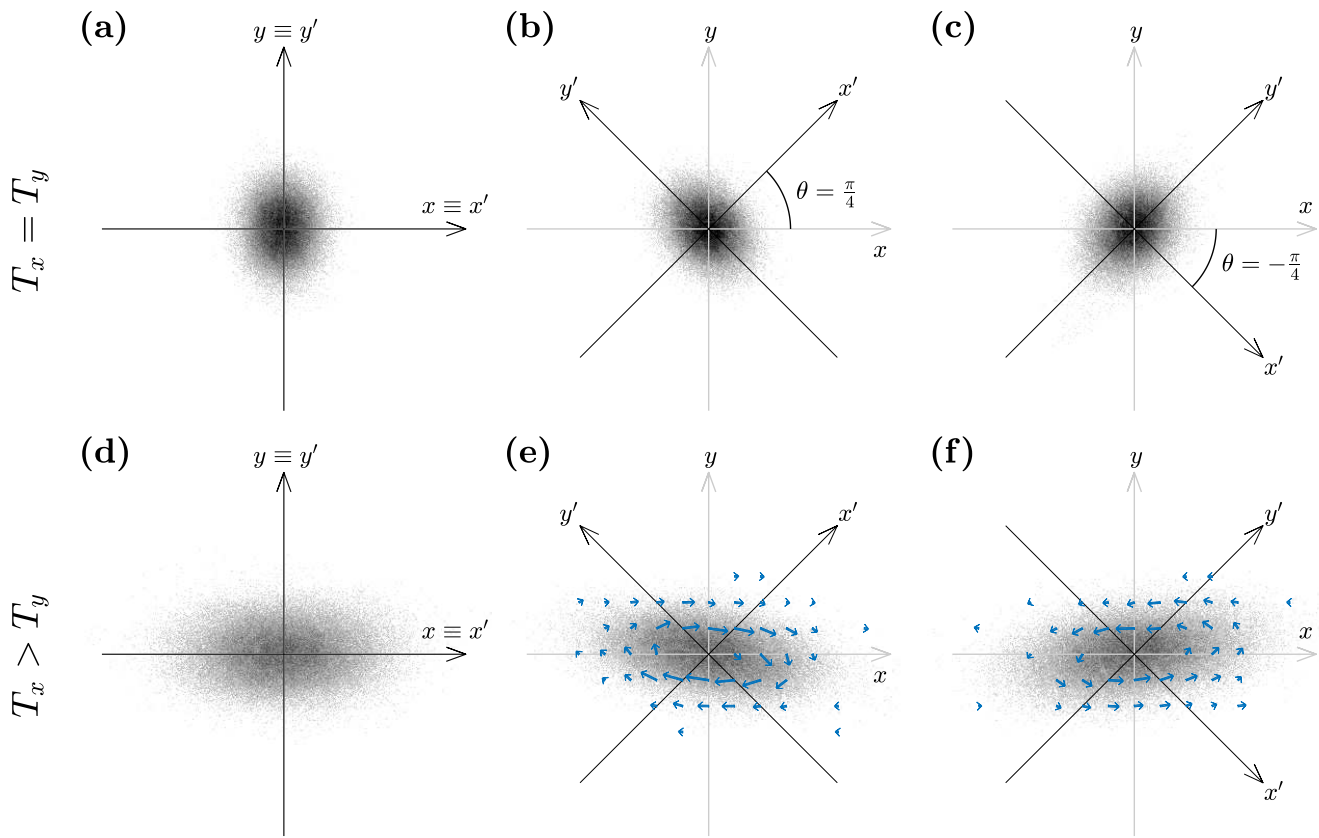}
\caption{
{\bf Brownian colloid in an elliptical potential.}
(a-c) Experimental steady-state probability distributions of a Brownian particle for isotropic temperature ($T_x=T_y=\unit[292]{K}$) inside an elliptical potential ($k_{x'}=\unit[1.63]{pN/\mu m}$, $k_{y'}=\unit[0.86]{pN/\mu m}$) (a) with its principal axes $x'$ and $y'$ aligned with the Cartesian coordinates $x$ and $y$ ($\theta=0$), (b) with $\theta=\pi/4$, and (c) with $\theta=-\pi/4$.
The probability densities are pictured by scatter plots of experimentally measured particle positions, darker regions corresponding to higher densities.
(d-f) Experimental steady-state probability distributions for anisotropic temperature ($T_x=\unit[1750]{K} > T_y=\unit[292]{K}$) (d) with $\theta=0$, (e) $\theta=\pi/4$, and (f) with $\theta=-\pi/4$.
(e-f) The blue arrows represent the associated probability flux (Eq.~\eqref{eq:v}): when the principle axes of the elliptical potential and the anisotropic thermal environment are rotated which respect to each other to break rotational symmetry, there appears a rotational flux component whose direction depends on the sign of $\theta$.
Note that there is no net flux when (a)-(c) the system is at thermal equilibrium, and when (d) the axes of the anisotropic temperature and the potential are aligned ($\theta=0$).
}
\label{fig1}
\end{figure*}

Theoretically, we model the motion of the Brownian particle using
overdamped Langevin equations in two dimensions \cite{Filliger2007Brownian}:
\begin{equation}
\label{eq:eom}
\left\{
\begin{array}{ccc}
\displaystyle \gamma \dot{x} 
& 
\displaystyle = 
&
\displaystyle -{\partial \over \partial x} U(x,y) + \sqrt{2 \gamma k_{\rm B} T_x} \, \xi_x(t)
\, , \\[3mm]
\displaystyle \gamma \dot{y} 
& 
\displaystyle = 
& 
\displaystyle -{\partial \over \partial y} U(x,y) + \sqrt{2 \gamma k_{\rm B} T_y} \, \xi_y(t)
\, .
\end{array}
\right.
\end{equation}
The particle's motion is confined by an elliptical harmonic potential $U(x,y)$ with stiffnesses $k_{x'}$ and $k_{y'}$ along its principal axes $x'$ and $y'$, which are rotated by an angle $\theta$ with respect
to the coordinate axes $x$ and $y$ (Fig.~\ref{fig1}):
\begin{equation}
U(x,y) = \frac{1}{2} \left[\begin{array}{cc}
x & y
\end{array}\right]
\,
{\bm R}(-\theta)
\,
{\bm k}
\,
{\bm R}(\theta)
\,
\left[\begin{array}{c}
x \\
y
\end{array}\right]
\, ,
\end{equation}
where
${\bm R}(\theta) 
= 
\left[\begin{array}{cc}
\cos \theta & \sin \theta \\ 
-\sin \theta & \cos \theta
\end{array}\right]$ 
and 
${\bm k}
= 
\left[\begin{array}{cc}
k_{x'} & 0 \\ 
0 & k_{y'}
\end{array}\right]$.
The coordinate axes $x$ and $y$ are aligned with the directions of the anisotropic temperatures $T_x$ and $T_y$, such that the angle $\theta$ provides a means to control the symmetry breaking between clockwise and counter-clockwise orientation. The corresponding thermal fluctuations are modeled by mutually independent Gaussian white noise sources $\xi_x(t)$ and $\xi_y(t)$ with $\langle \xi_x(t) \rangle = \langle \xi_y(t) \rangle = 0$ and $\langle \xi_x(t)\xi_x(t') \rangle = \langle \xi_y(t)\xi_y(t') \rangle = \delta(t-t')$.
In the following, $T_y$ is equal to the temperature of the aqueous solution, i.e. room temperature $T_y=\unit[292]{K}$, while $T_x$ is either room temperature or hotter due to the effective heating from the electric noise signals \cite{martinez2013effective,mestres2014realization,dinis2016thermodynamics}. 
The viscous friction in Eq.~\eqref{eq:eom} is given by the isotropic Stokes coefficient $\gamma=6\pi\nu R$, where $\nu$ is the viscosity of the watery solution and $R$ the particle radius. 

Experimentally, we use polystyrene particles with diameter $2R =1.98\,{\rm \mu m}$ (Microparticles GmbH) held in a potential generated using an optical tweezers \cite{jones2015optical}: we focus a laser beam (wavelength $\lambda = 532\,{\rm nm}$) using a high-numerical aperture objective ($60\times$, NA 1.40), while we introduce the ellipticity in the potential by altering the intensity profile of the laser beam using a spatial light modulator (PLUTO-VIS, Holoeye GmbH). We track the position of the particle at $400\,{\rm fps}$ by digital video microscopy using the radial symmetry algorithm \cite{parthasarathy2012rapid}. The values of the optical trapping stiffnesses, $k_{x'}=\unit[1.63]{pN/\mu m}$ and $k_{y'}=\unit[0.86]{pN/\mu m}$, are measured from the acquired particle trajectories by using the equipartition method and the autocorrelation methods \cite{jones2015optical} in the absence of electric noise (i.e. when $T_x=T_y=\unit[292]{K}$). The 
Figures~\ref{fig1}(a)-(c) show the experimental equilibrium probability density $p_{\mathrm{ss}}(x,y)$ of the particle in the elliptical trap with $\theta = 0$ (Fig.~\ref{fig1}(a)), $\theta=\pi/4$ (Fig.~\ref{fig1}(b)), and $\theta=-\pi/4$ (Fig.~\ref{fig1}(c)), when $T_x = T_y = \unit[292]{K}$ are both equal to room temperature. 

We can now establish a non-equilibrium steady state by introducing different temperatures along the $x$- and $y$-directions. Due to the colloidal particle being electrically charged in solution, a randomly oscillating field applied along the $x$-direction produces a fluctuating electrophoretic force on the particle, which increases its random fluctuations along the $x$-direction, leading to an effective increase of the temperature \cite{martinez2013effective}. The electric field is generated by driving with an electric white noise two parallel thin wires (gold, diameter $30\,{\rm \mu m}$) placed on either side of the optical trap at a distance of $1\,{\rm mm}$.
The effective temperature along $x$ is then proportional to the variance of the particle position along the $x$-direction when the principal axes of the optical trap are aligned with the Cartesian axes $x$ and $y$.
Figures~\ref{fig1}(d)-(f) present the resulting stationary probability distributions for the cases $\theta = 0$ (Fig.~\ref{fig1}(d)), $\theta=\pi/4$ (Fig.~\ref{fig1}(e)), and $\theta=-\pi/4$ (Fig.~\ref{fig1}(f)), when $T_x =\unit[1750]{K}$ and $T_y = \unit[292]{K}$: they are elongated along the $x$-direction (in comparison with Figs.~\ref{fig1}(a)-(c)), because of the presence of the extra noise. 
Furthermore, we can measure the stationary probability density current according to \cite{seifert2012stochastic}
\begin{multline}
\label{eq:v}
\left[
\begin{array}{c}
J_x(x,y) \\�J_y(x,y)
\end{array}
\right]
=
\left[
\left\langle
\begin{array}{c}
x(t+\Delta t)-x(t) \\ y(t+\Delta t)-y(t)
\end{array}
\right\rangle_{x(t)=x, \, y(t)=y}
\right.
\\
+ \left.
\left\langle
\begin{array}{c}
x(t)-x(t-\Delta t) \\ y(t)-y(t-\Delta t)
\end{array}
\right\rangle_{x(t)=x, \, y(t)=y}
\right]
\frac{p_{\mathrm{ss}}(x,y)}{2\Delta t}
\, ,
\end{multline}
where the averages are taken over all particle displacements during a sampling time interval $\Delta t$, which start (first line) or end (second line) at position $\left[\begin{array}{c} x \\�y \end{array}\right]$. This current is represented by the blue arrows in Figs.~\ref{fig1}(e)-(f) and clearly indicates the presence of a gyrating motion; the strength and the direction of this rotational motion depend on the rotary asymmetry induced by $\theta$ and, importantly, they vanish for $\theta=0$ (Fig.~\ref{fig1}(d)), because the principal directions of the baths and of the potential are aligned and therefore there is no symmetry breaking.
Note also that there is essentially no flux in Figs.~\ref{fig1}(a)-(c), as expected at thermal equilibrium.

\begin{figure*} [ht!]
\includegraphics[width=\textwidth]{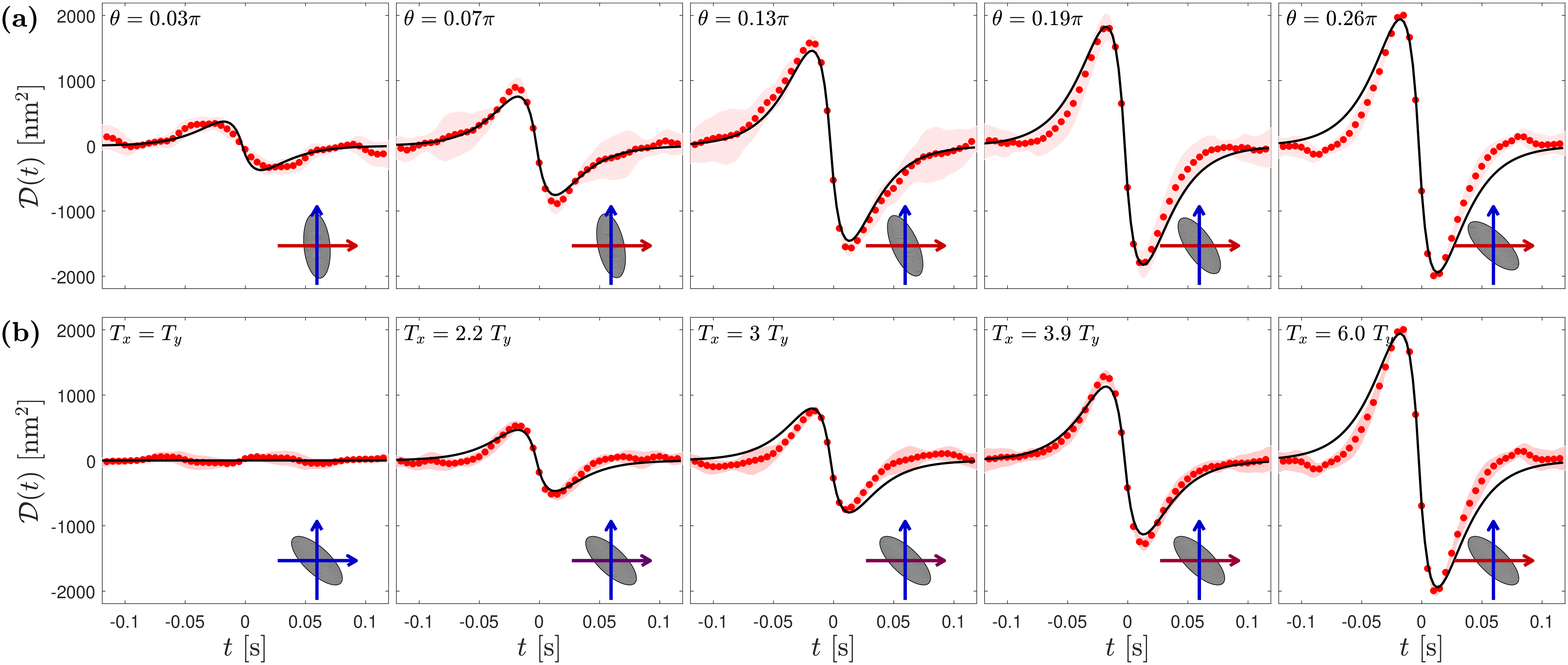}
\caption{
{\bf Cross-correlation functions $\mathcal{D}(t)$.} 
(a) $\mathcal{D}(t)$ as a function of the relative orientation $\theta$ between the axes of the temperature anisotropy ($T_x \equiv 6.0\,T_y$) and those of the potential; it is maximized for $\theta=\pi/4$. 
(b) $\mathcal{D}(t)$ as a function of the temperature anisotropy ($\theta \equiv \pi/4$); it grows with increasing $T_x$.  The red symbols represent the experimental data (the shaded area is the standard deviation) and the solid black lines represent the corresponding theory (Eq.~\eqref{dCCF}). The insets show schematically the alignment between the axes of the temperature anisotropy (arrows) and those of the potential (ellipses), the temperature along the $y$-axis is color-coded in blue indicating the ``cold'' direction
$T_y=\unit[292]{K}$, while the temperature in $x$-direction is indicated in blue
if $T_x=T_y=\unit[292]{K}$ and red if it corresponds to the ``hot'' direction with $T_x>T_y$.
}
\label{fig2}
\end{figure*}

In order to quantify this rotational behavior, we calculate the
differential cross correlation function between $x$ and $y$ \cite{volpe2006torque,volpe2007brownian}: 
\begin{align}
\mathcal{D}(t) &= 
\langle x(t^\ast)y(t^\ast+t) \rangle - \langle y(t^\ast)x(t^\ast+t) \rangle
\nonumber \\
& =
\langle r(t^\ast) r(t^\ast+t) \sin( \phi(t^\ast+t) - \phi(t^\ast) ) \rangle
\, , 
\label{eq:D}
\end{align}
where the angular brackets indicate the average over the steady-state distribution,
for which $\mathcal{D}(t)$ is independent of the
reference time point $t^\ast$.
Its representation in the second line using polar coordinates
$r = \sqrt{x^2+y^2}$ and $\phi=\arctan(y/x)$ 
illustrates that it vanishes
if there is no net motion of the colloid and that it is positive (negative)
for counter-clockwise (clockwise) net gyrating movements.

Since the model described by Eq.~\eqref{eq:eom} can be solved analytically,
we can calculate an exact closed expression for $\mathcal{D}(t)$
(see Appendix \ref{app:math} for details),
\begin{equation}\label{dCCF}
\mathcal{D}(t) = {\mathrm{sign}}(t) \,
k_{\rm B} (T_x-T_y) \,
\frac{ e^{-\frac{|t|k_{x'}}{\gamma}} -e^{-\frac{|t|k_{y'}}{\gamma}} }{k_{x'}+k_{y'}} \,
\sin(2\theta)
\, .
\end{equation}
Experimentally, $\mathcal{D}(t)$ can be directly evaluated from the recorded trajectories without explicit knowledge of the trap parameters and temperatures \cite{volpe2006torque,volpe2007brownian}, using the expression in Eq.~\eqref{eq:D}.
Figure~\ref{fig2} presents the experimental $\mathcal{D}(t)$ (red symbols) for different values of $\theta$ and temperature anisotropy, which are in good agreement with the theoretical predictions from Eq.~\eqref{dCCF}
(black lines): $\mathcal{D}(t)$ vanishes when $\theta=0$ and is maximized when $\theta = \pm \pi/4$ (Fig.~\ref{fig2}(a)); and $\mathcal{D}(t)$ increases as the temperature anisotropy increases (Fig.~\ref{fig2}(b)).

\begin{figure}[t]
\includegraphics[width=.5\textwidth]{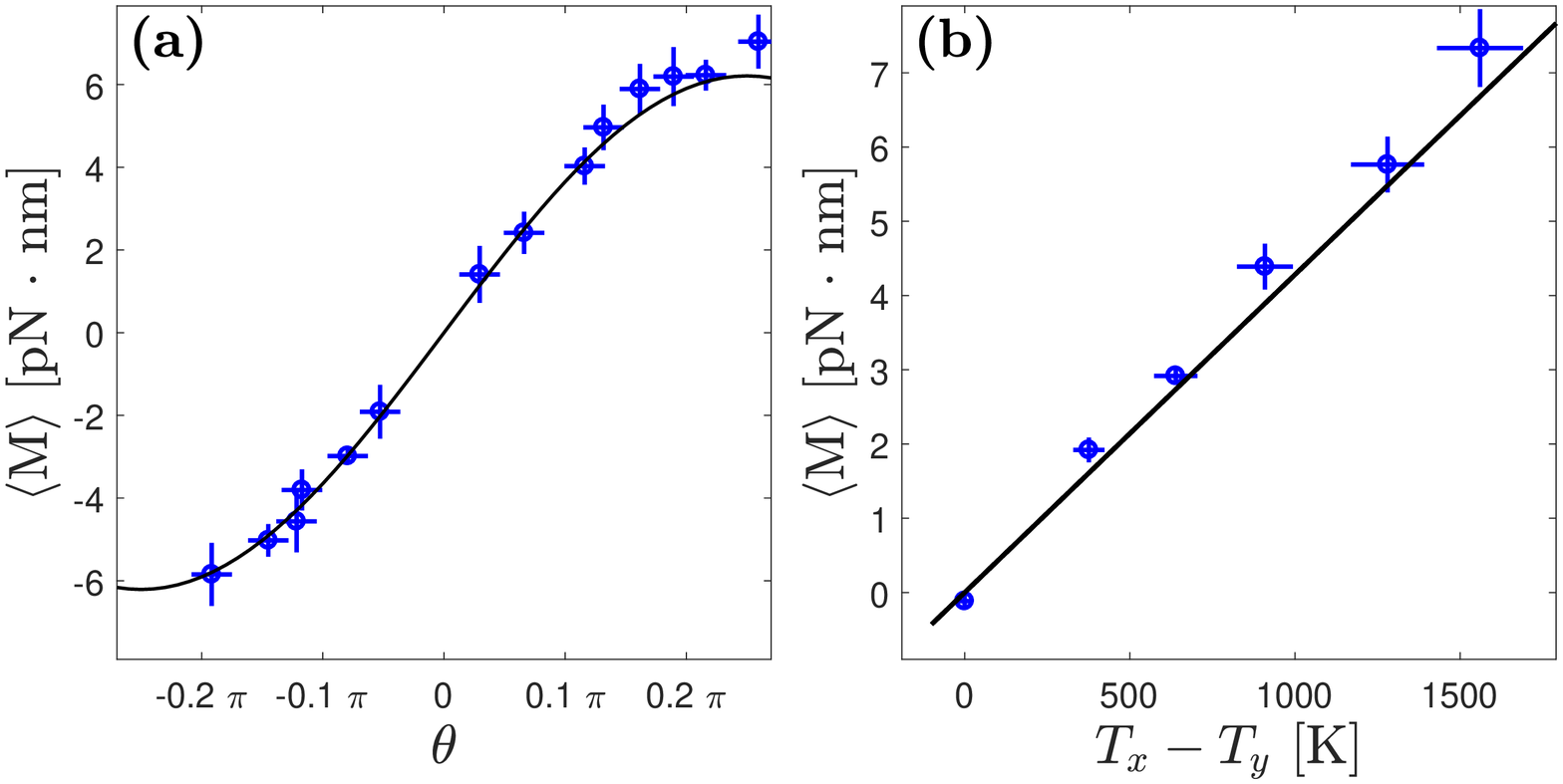}
\caption{
{\bf Torque} as a function of (a) the relative orientation
$\theta$ between the axes of the temperature anisotropy ($T_x \equiv 6.0\,T_y$)
and those of the potential, and (b) as a function of the temperature difference $T_x-T_y$
($\theta \equiv \pi/4$). The symbols represent the experimental data
(corresponding to 5 trajectories of $\unit[50]{s}$, the error bars indicate
standard deviations) and the solid lines are the theoretical predictions given by Eq.~\eqref{rotat}.
}
\label{fig3}
\end{figure}

\begin{figure}[t]
\includegraphics[width=.5\textwidth]{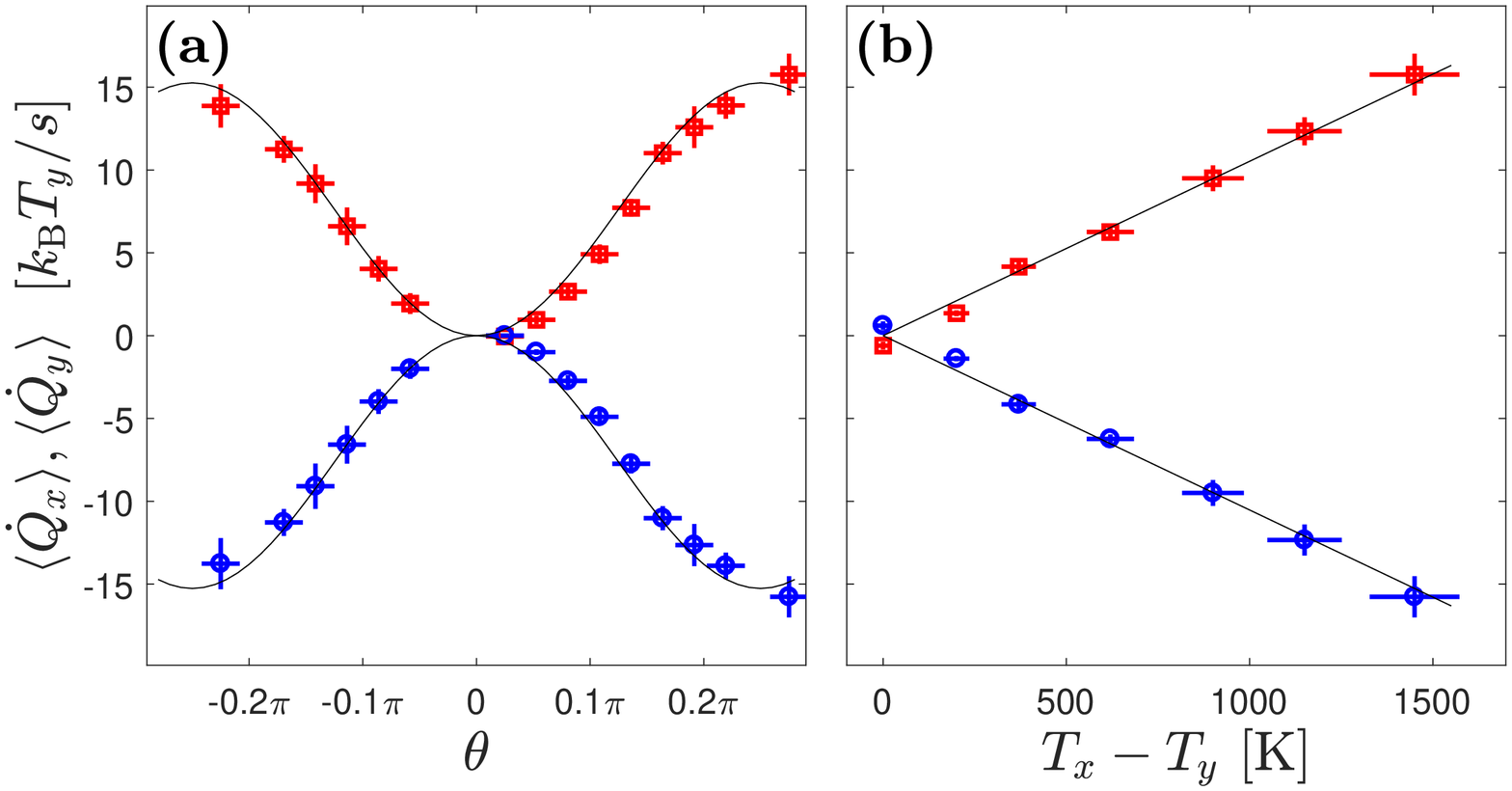}
\caption{
{\bf Heat flow between the two baths} as a function of (a) the relative orientation $\theta$ between the axes of the temperature anisotropy ($T_x=6.0\, T_y$) and those of the potential, and (b) as a function of the temperature difference $T_x-T_y$ ($\theta=\pi/4$).
The red squares and blue circles represent the experimental data measuring $\dot{Q}_x$ and $\dot{Q}_y$, respectively (corresponding to 5 trajectories of $\unit[50]{s}$, the error bars indicate standard deviations) and the solid lines represent the theoretical predictions given by Eqs.~\eqref{heatx} and \eqref{heaty}.
}
\label{fig4}
\end{figure}

The rotational motion of the particle around the origin can also be measured by studying its weighted angular velocity $r^2 \mathrm{d}\phi/\mathrm{d}t$. Its average is proportional to the strength $M$ of the average torque
$\left\langle x(\partial U/\partial y) - y(\partial U/\partial x) \right\rangle$
exerted by the particle on the potential $U$ \cite{volpe2006torque,volpe2007brownian,Filliger2007Brownian}:
\begin{equation}\label{eqieoosdn}
M =- \gamma \left\langle r^2\frac{d\phi}{dt} \right\rangle
\, ,
\end{equation}
This expression provides a way of computing the average torque directly from the recorded trajectories without explicit knowledge of the trap parameters and temperatures.
On the other hand, an analytical prediction for the torque as a function of precisely these parameters is again obtained from the exact solution of Eq.~\eqref{eq:eom} \cite{Filliger2007Brownian} (see Appendix \ref{app:math} for details),
\begin{equation}
\label{rotat}
\gamma \left\langle r^2\frac{\mathrm{d}\phi}{\mathrm{d}t} \right\rangle
= -M 
= -k_{\rm B}(T_x-T_y) \frac{k_{x'}-k_{y'}}{k_{x'}+k_{y'}} \sin(2\theta)
\, .
\end{equation}
By the independent measurement of $k_{x'}$, $k_{y'}$, $T_x$, and $T_y$, we can compare this prediction with the experimental measurements without any fit parameter. The symbols in Fig.~\ref{fig3} represent the experimentally measured torques, which are indeed in very good agreement with theoretical predictions from Eq.~\eqref{rotat} (solid lines). 
When evaluating the torque from the experimental data, we used an estimator which is exact to first order in the sampling time step (as opposed to the zeroth order naive estimator) in order to obtain an accurate value despite the relatively large experimental value $\Delta t = \unit[2.5]{ms}$ (see the Appendix \ref{app:Dt} for details).
The torque vanishes when $\theta = 0$ (Fig.~\ref{fig3}(a)) and when $T_x = T_y$ (Fig.~\ref{fig3}(b)), increases as $\theta$ approaches $\pi/4$ and grows linearly with the temperature difference $T_x-T_y$. 

The presence of a systematic rotational motion of the particle is connected to a transfer of heat from the hot to the cold bath. 
Following Sekimoto's stochastic energetics approach \cite{sekimoto2010stochastic},
we identify heat with the work performed by the dissipating and thermally fluctuating
forces, so that the heat absorbed by the particle from the hot reservoir at
temperature $T_x$
along the trajectory $[x(t), y(t)]$ reads
\begin{align}
\label{eq:DefExchangedHeat}
Q_x(\tau) = \int_0^\tau \left[ -\gamma \dot x(t) + \sqrt{2 k_{\rm B} T_x \gamma} \, \xi_x(t) \right] \circ \mathrm{d}x(t)
\, ,
\end{align}
where $\circ$ denotes the Stratonovich product.
Using the equations of motion \eqref{eq:eom}, this can be rewritten as
\begin{equation}
\label{eq:ExchangedHeatFromPotential}
Q_x(\tau) = \int_0^\tau {\partial \over \partial x} U(x(t),y(t)) \circ \mathrm{d}x(t) .
\end{equation}
This equation expresses the heat flow from the hot reservoir to the colloidal particle entirely by means of experimentally accessible quantities, i.e. $x(t)$, $y(t)$, $\gamma$, $k_{x'}$, $k_{y'}$, and $\theta$. In the stationary state, the average heat absorbed along trajectories divided by the observation time $\langle\dot{Q}_x\rangle = \langle Q_x(\tau) \rangle /\tau$ is a constant independent of the length $\tau$ of the trajectory.
This average heat flow can be calculated analytically as (see Appendix \ref{app:math} for derivation)
\begin{equation}
\label{heatx}
\langle \dot{Q}_x \rangle = \frac{k_{\rm B}(T_x-T_y)}{4\gamma(k_{x'}+k_{y'})} \left[(k_{x'}-k_{y'})\sin(2\theta) \right]^2
\, .
\end{equation}
An analogous formula holds for the heat absorbed from the cold reservoir at temperature $T_y$,
\begin{equation}
\label{heaty}
\langle \dot{Q}_y \rangle =-\langle \dot{Q}_x \rangle
\, .
\end{equation}

In Fig.~\ref{fig4}, we present the experimentally measured heat flows from the cold reservoir (blue circles) and from the hot reservoir (red squares) to the particle as a function of $\theta$ (Fig.~\ref{fig4}a) and $T_x-T_y$ (Fig.~\ref{fig4}b).
As in the case of the torque, also when evaluating the heat flow from the experimental data according to Eq.~\eqref{eq:ExchangedHeatFromPotential}, as in the case of the torque, we used an estimator which is accurate to first order in the sampling time step $\Delta t$ (see Appendix \ref{app:Dt} for details).
These experimental results are in very good agreement with the theoretical predictions \eqref{heatx} and \eqref{heaty} (solid lines). The average direction of the heat flow is always from the hot to the cold reservoir; its intensity vanishes as $\theta \to 0$ and $T_x \to T_y$, and increases as $\theta \to \pi/4$ and as $T_x$ increases. In the current setup this heat flow is turned into systematic motion, but is not used to perform work against an external load, such that efficiency as the ratio between work performed and heat taken up from the hotter reservoir cannot be defined.

In conclusion, we have presented an experimental realization of a microscopic heat engine employing a single colloidal particle moving in a generic elliptical optical trap while in simultaneous contact with two heat reservoirs. This experimental model features a minimal degree of complexity necessary to obtain a microscopic, circularly operating heat engine generating a torque from which work can in principle be extracted \cite{Filliger2007Brownian}. Furthermore, it has the advantage of being completely solvable analytically, therefore providing an ideal testbed to compare theory and experiments. 

We finally point out a very recent interesting experimental work \cite{chiang2017electrical}, which studies a physically completely different but mathematically equivalent system, namely two capacitively coupled resistor-capacitor circuits whose dynamical equations for the two voltages can be mapped to the model described by Eq.~\eqref{eq:eom}.

\begin{acknowledgments}
All authors aknowledge useful discussion with the members of Yellow Thermodynamics and with Jan Wehr.
This work was partially supported by the European Research Council ERC Starting Grant ComplexSwimmers (grant number 677511) and by the Marie Sklodowska-Curie Individual Fellowship ActiveMotion3D (grant number 745823).
RE and SB acknowledge financial support from the Swedish Research Council (Vetenskapsr{\aa}det) under the grants No.~621-2013-3956, No.~638-2013-9243 and No.~2016-05412. 
LD acknowledges financial support by the Stiftung der Deutschen Wirtschaft.
\end{acknowledgments}

\appendix

\section{Solution of the model}
\label{app:math}

\subsection{Dynamics}

Starting from the Eqs.~\eqref{eq:eom} and compactifying notation, the overdamped equations of motion read:
\begin{equation}\label{eq:ModelLangevinOverdamped}
	\dot{\bm r}(t) = -\bm A \bm r(t) + \bm B \bm\xi(t) \, ,
\end{equation}
with $\bm{r} = \begin{bmatrix} r_1 \\ r_2 \end{bmatrix} \equiv \begin{bmatrix} x \\ y \end{bmatrix}$,
$\bm{\xi}(t) = \begin{bmatrix} \xi_1(t) \\ \xi_2(t) \end{bmatrix} \equiv \begin{bmatrix} \xi_x(t) \\ \xi_y(t) \end{bmatrix}$,
\begin{equation}\label{eq:explicitA}
{\bm A}
=
\frac{1}{\gamma} 
{\bm R}(-\theta)
\,
{\bm k}
\,
{\bm R}(\theta)
\, ,
\end{equation}
and
\begin{equation}
{\bm B}
=
\begin{bmatrix}
\sqrt{2 k_{\rm B}T_x / \gamma}  & 0 \\ 0 & \sqrt{2\gamma k_{\rm B}T_y / \gamma} 
\end{bmatrix}
\, .
\end{equation}
Equation \eqref{eq:ModelLangevinOverdamped} is the stochastic differential equation (SDE) of a general Ornstein-Uhlenbeck process and is equivalent to a Fokker-Planck equation \cite{Risken:FokkerPlanckEquation, Gardiner:1985} for the transition probabilities, or propagator, $p(t, \bm r | t_0,  \bm{r_0})$,
\begin{align}
	\partial_t p(t, \bm{r} | t_0,  \bm{r_0}) = \sum_{i,j}  A_{ij} \partial_{r_i} \left[ r_j \, p(t, \bm{r} | t_0,  \bm{r_0}) \right] \nonumber \\ + D_{ij} \partial_{r_i}\partial_{r_j}p(t, \bm{r} | t_0,  \bm{r_0}) 
	\label{eq:ModelFPOU}
\end{align}
with the diffusion matrix
\begin{align}
	{\bm D}=\frac{1}{2} {\bm B} {\bm B}^{\rm T}
	=
	\begin{bmatrix}
		k_{\rm B} T_x/\gamma & 0 \\ 0 & k_{\rm B} T_y/\gamma
	\end{bmatrix}.
\end{align} 
The propagator gives the probability to find the particle in an infinitesimal volume element $\mathrm d^2 r$ around $\bm r$ at time $t$ given that it was at $\bm r_0$ at an earlier time $t_0$. Since the system is Markovian, its statistics are fully determined by $p$ and some initial distribution $p_0(\bm r_0)$. The Fokker-Planck equation~\eqref{eq:ModelFPOU} can be solved exactly \cite{Risken:FokkerPlanckEquation, Gardiner:1985} and the resulting propagator is
\begin{align}\label{eq:OUFPPropagator}
	p(t, \bm{r} | t_0, \bm{r}_0) = \frac{\mathrm{e}^{-\frac{1}{2} \, {\left[ \bm{r} - e^{-(t-t_0) \bm A} \bm r_0 \right]}^T \bm{\Sigma}^{-1}(t-t_0) \left[ \bm r - e^{-(t-t_0) \bm A} \bm r_0 \right] }}{\sqrt{(2\pi)^2 \, \det \bm\Sigma(t-t_0)}} ,
\end{align}
where the covariance matrix is
\begin{align}\label{eq:OUFPSigma}
	{\bm \Sigma}(t) = {\bm \Sigma}(\infty) - e^{-t {\bm A}} {\bm \Sigma}(\infty) e^{-t {\bm A}^{\rm T}} 
\end{align}
and ${\bm \Sigma}(\infty)$ is obtained as the solution of the matrix equation
\begin{align}
	{\bm A} {\bf \Sigma}(\infty) + {\bm \Sigma}(\infty) {\bm A}^{\rm T} = 2 {\bm D}.
\end{align}
One can see that $\bm r(t)$ is a time-homogeneous Gaussian process. For our system, we find
\begin{multline}\label{eq:OUFPSigmaPart}
	{\bm \Sigma}(\infty) = \frac{1}{\operatorname{tr} {\bm A}  \, \operatorname{det}{\bm A}}
	\left[
	\begin{array}{c}
	D_2 \, A_{12}^2 + D_1 (A_{22}^2 + \det{\bm A}) \\
		-D_1 \, A_{21} \, A_{22} - D_2 \, A_{11} \, A_{12}
	\end{array}
	\right.
	\\
	\left.
	\begin{array}{c}
	-D_1 \, A_{21} \, A_{22} - D_2 \, A_{11} \, A_{12}   \\
	 D_1 \, A_{21}^2 + D_2 (A_{11}^2 + \det{\bm A}) 
	\end{array}
	\right]
\end{multline}
where $D_i = k_{\rm B} T_i / \gamma$ are the diagonal entries of the matrix ${\bm D}$, $\det{\bm A}=k_{x'}k_{y'}/\gamma^2$ and ${\rm Tr}({\bm A})=(k_{x'}+k_{y'})/\gamma$.

\subsection{Steady state}

From the solution~\eqref{eq:OUFPPropagator} and the positive definiteness of $\bm A$, we understand that the system reaches a steady state in the limit $t \to \infty$, whose distribution is
\begin{equation}\label{eq:OUFPSteadyState}
p_{\mathrm{ss}}(\bm r) = \frac{e^{ -\frac{1}{2} \, { {\bm r}^{\rm T} } {\bm \Sigma}^{-1}(\infty) {\bm r} } }{\sqrt{(2\pi)^2 \, \det {\bm \Sigma}(\infty)}} \, .
\end{equation}
The characteristic relaxation time to the steady state is $\tau_\infty = (\det {\bm A})^{-1/2} = \gamma/\sqrt{ k_{x'} k_{y'} }$, which is approximately $\unit[16]{ms}$ for our experiments.

\subsection{Correlations}

At the steady state, the
autocorrelation matrix ${\bm{\mathcal{C}}}(t)$ (with entries $\mathcal{C}_{ij}(t)=\langle  r_i(t^\ast+t) r_j(t^\ast)\rangle$) becomes independent of the reference time $t^\ast$, and, using~\eqref{eq:OUFPPropagator},
can be computed as
\begin{align} \label{eq:AutocorrelationMatrix}
	\displaystyle
	{\bm{\mathcal{C}}}(t)= \begin{cases} 
		e^{-t {\bm A}}{\bf \Sigma}(\infty) & t\ge 0 \\
		{\bm \Sigma}(\infty)e^{-|t| {\bm A}^{\rm T}}   & t<0    \end{cases}.
\end{align}
A measure of the non-equilibrium state of the system and the particle's rotational motion is the asymmetry in the correlation function, i.e., the differential cross correlation function:
\begin{align}
	\mathcal{D}(t)&=	
	\langle x(t^\ast)y(t^\ast+t)\rangle-\langle y(t^\ast)x(t^\ast+t)\rangle
	\nonumber\\
	&= \langle r_1(t^\ast)r_2(t^\ast+t)\rangle-\langle r_2(t^\ast)r_1(t^\ast+t)\rangle
	\nonumber\\
	&= \mathcal{C}_{21}(t)-\mathcal{C}_{12}(t) \, .
\end{align}
Plugging in~\eqref{eq:AutocorrelationMatrix}, we obtain formula~\eqref{dCCF} above.

\subsection{Torque}

The rotational motion of the particle around the origin can also be assessed by studying its weighted angular velocity $r^2 d\phi/dt$, a quantity reminiscent of angular momentum.
Using the equations of motion~\eqref{eq:eom}, we find that its average $\langle r^2 d\phi/dt \rangle = \langle x \dot{y} - y \dot{x}\rangle$ is related to the average torque $M$ exerted on the potential (introduced by Filliger and Reimann \cite{Filliger2007Brownian} and in Eq.~\eqref{eqieoosdn} above):
\begin{align} \displaystyle
	-\gamma \left\langle r^2 \frac{d \phi}{dt} \right\rangle
	&= M = \left\langle x \frac{\partial U}{\partial y} - y \frac{\partial U}{\partial x}
	\right\rangle \nonumber \\
	&= \frac{k_B (T_x - T_y) (k_{x'} - k_{y'}) \sin(2\theta)}{k_{x'} + k_{y'} }.
\end{align}
The left-hand representation provides a way of computing the average torque directly from the trajectories and independently of the trap parameters and temperatures.

\subsection{Heat absorbed along a trajectory}

The equations of motion (\ref{eq:ModelLangevinOverdamped}) specify several forces acting on the particle, each of which can be associated with a physical component of the system. In particular, we have the potential force ${\bm F}^{U} = -{\bm \nabla} U$ as well as forces linked to the interaction with the medium and reservoirs, namely the frictional force ${\bm F}^{\rm diss} = -{\gamma} \dot {\bm r}$ and the thermal fluctuations ${\bm F}^{\text{therm}} = \gamma\bm B{\bm \xi}(t)$. The equations of motion merely state the balance of these forces.

Following Sekimoto's stochastic energetics approach, we identify heat with the work performed by the dissipating and thermally fluctuating forces so that the heat absorbed by the particle can be
calculated using Eqs.~\eqref{eq:DefExchangedHeat} and \eqref{eq:ExchangedHeatFromPotential}.
Explicitly, the heat flowing from the reservoirs to the system along a trajectory $\bm r(t)$ reads
\begin{align} \label{eq:ExchangedHeatFromPotential_explicit1}
Q_x &= \int_0^\tau \left[ -\gamma \dot x(t) + \sqrt{2 \gamma k_{\mathrm B} T_x} \xi_x(t) \right] \circ \mathrm{d}x(t)
	\notag \\
	&= \left(k_{x'}\cos(\theta)^2+k_{y'}\sin(\theta)^2\right)\frac{x^2(\tau)-x^2(0)}{2} 
	\notag \\
	&\quad\mbox{}+(k_{x'}-k_{y'})\sin(\theta)\cos(\theta)\int_0^\tau y(t) \circ \mathrm{d}x(t) 
\end{align}
and
\begin{align}
Q_y &= \int_0^\tau \left[ -\gamma \dot y(t) + \sqrt{2 \gamma k_{\mathrm B} T_y} \xi_y(t) \right] \circ \mathrm{d}y(t)
	\notag \\
	&= \left(k_{x'}\sin(\theta)^2+k_{y'}\cos(\theta)^2\right)\frac{y^2(\tau)-y^2(0)}{2}
	\notag \\
	&\quad\mbox{}+(k_{x'}-k_{y'})\sin(\theta)\cos(\theta)\int_0^\tau x(t) \circ \mathrm{d}y(t)
	\label{eq:ExchangedHeatFromPotential_explicit2} \, ,
\end{align}
where the second equality in both relations follows using the equations of motion~\eqref{eq:eom}.
These relations form the basis for evaluating the average heat flow in Fig.~\ref{fig4} from the experimental data.
Evaluating the integrals along any trajectory in the stationary state as averages
over the steady-state distribution, we obtain Eqs.~\eqref{heatx} and~\eqref{heaty}.

\section{Finite-time estimators}
\label{app:Dt}

\begin{figure}[t]
\includegraphics[width=.5 \textwidth]{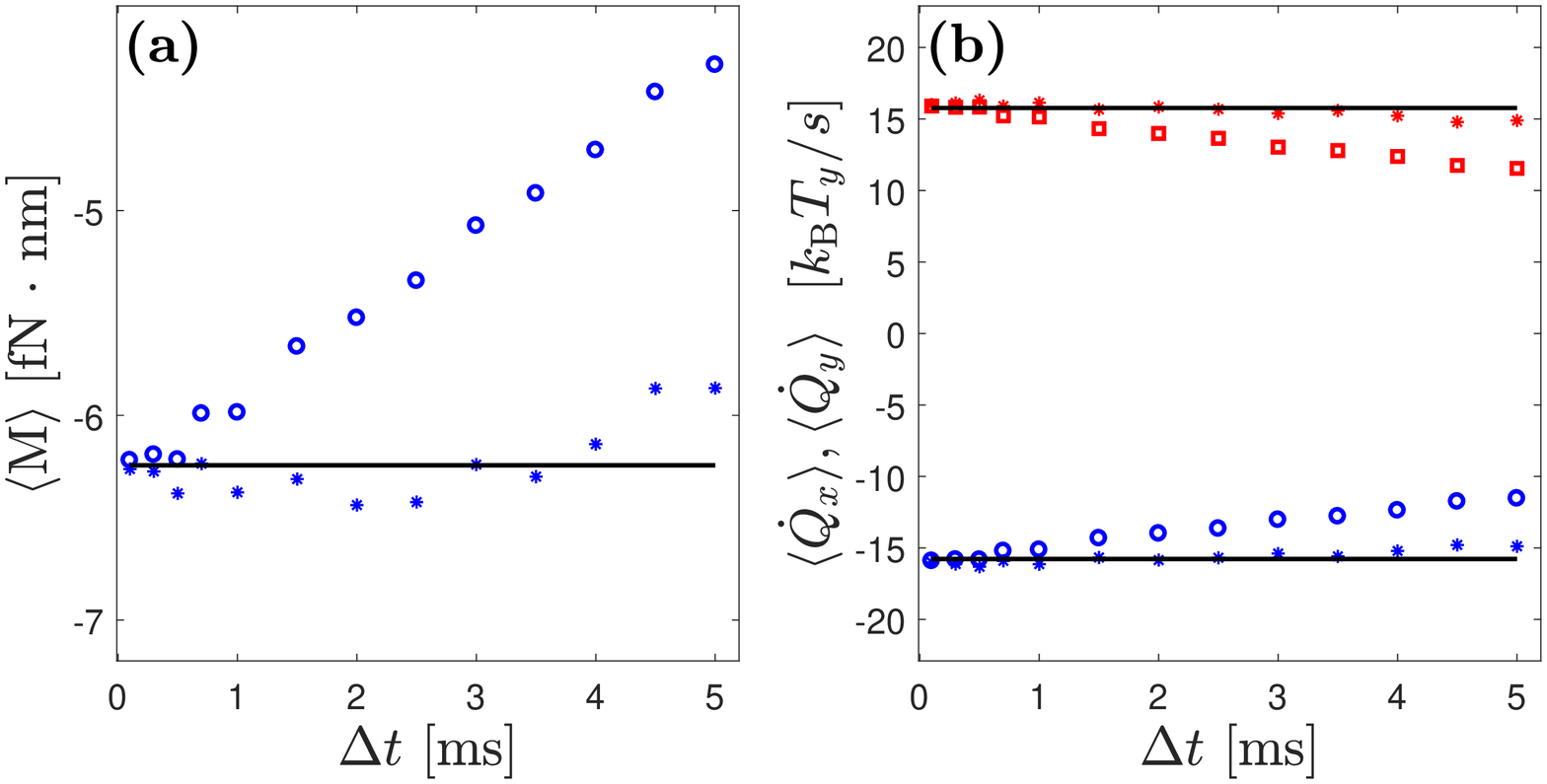}
\caption{
{\bf Estimation of torque and the heat flow.} 
Simulations showing the measured values (a) of the torque and (b) of the heat flows as a function of the sampling time step $\Delta t$ using a zeroth-order estimator (circles) and a first-order estimator (Eq.~\eqref{eq:estimatorf}, crosses). The black lines represent the exact values predicted by the theory. The first-order estimator permits us to obtain the correct values at the experimental sampling time step $\Delta t = \unit[2.5]{ms}$.
}
\label{fig5}
\end{figure}

The estimation of the torque and heat flows from the experimental data using Eqs.~\eqref{eqieoosdn} and \eqref{eq:ExchangedHeatFromPotential}, respectively, depends sensitively on the sampling time step $\Delta t$ of the recorded trajectories.
More precisely, the estimators for the time-averaged quantities are discretizations of the integrals
\begin{equation}
	M \simeq -\frac{\gamma}{\tau} \int_0^\tau r^2(t) \circ \mathrm{d}\phi(t) 
	\quad\text{and}\quad
	\langle \dot Q_{x,y} \rangle \simeq \frac{Q_{x,y}}{\tau}
\end{equation}
with $Q_{x,y}$ given in Eqs.~\eqref{eq:ExchangedHeatFromPotential_explicit1}
and~\eqref{eq:ExchangedHeatFromPotential_explicit2}, respectively. Let us denote the estimators for sampling time step $\Delta t$ by $M(\Delta t)$ and $\langle \dot Q_{x,y} \rangle(\Delta t)$. The exact (experimental) values are obtained in the limit $\Delta t \to 0$. However, $\Delta t$ is subject to experimental constraints and in the case of our experiments is $\Delta t =\unit[2.5]{ms}$ (corresponding to $\unit[400]{fps}$), which is not sufficiently small to obtain accurate values using the naive estimators $M(\Delta t)$ or $\langle \dot Q_{x,y} \rangle(\Delta t)$. To illustrate this, we performed simulations of the system under the same experimental conditions \cite{volpe2013simulation} and computed the resulting torque and heat flows for different sampling time steps $\Delta t$ as shown by the open circles and squares in Fig.~\ref{fig5}. For too large sampling times, these estimates clearly deviate from the analytic predictions (solid lines).

In order to solve this problem, we introduce an improved, first-order estimator, which we will explain using the example of the torque; the method works completely analogously for the heat flows. We first note that the torque (as well as the heat flows) in the steady state are constant in time, such that any dependencies on the sampling times step $\Delta t$ (as the ones detected when estimating torque and heat flows from the simulations shown in Fig.~\ref{fig5}) are due to the simple estimators $M(\Delta t)$ (and $\langle \dot Q_{x,y} \rangle(\Delta t)$) being too imprecise. 
Assuming that $M(\Delta t)$ is a smooth function of the sampling time step, we can expand it around the exact value $M(0)$,
\begin{equation}\label{eq:estimator1}
M(\Delta t) = M(0)  + b \Delta t + \mathcal{O}(\Delta t^2)
\, ,
\end{equation} 
where $b$ is the linear deviation coefficient and $\mathcal{O}(\Delta t^2)$ stands for higher order deviations. Likewise, the average torque sampled for a time step of size $2 \Delta t$ is
\begin{equation}\label{eq:estimator2}
M(2 \Delta t) = M(0) + 2 b \Delta t + \mathcal{O}(\Delta t^2)
\, .
\end{equation} 
Combining these two expressions, we can eliminate the linear deviation $b$, so that the exact value $M(0)$ can be approximated as
\begin{equation}
	M(0) =
\underbrace{M(\Delta t)}_{\mbox{zeroth order}}
+ 
\underbrace{\left[ M(\Delta t - M(2\Delta t)\right]}_{\mbox{first-order correction}}
+ \mathcal{O}(\Delta t^2) \, ,
\end{equation}
up to second-order deviations in $\Delta t$.
In other words, we construct an improved estimator
\begin{equation}
\label{eq:estimatorf}
M^\ast(\Delta t) = 2M(\Delta t) - M(2\Delta t)
\end{equation}
from the naive estimator $M(\Delta t)$, which is accurate to first order in $\Delta t$ as opposed to the zeroth order precision of $M(\Delta t)$. In principle, we could continue this scheme to higher orders by including measurements of $M$ at higher multiples of  the sampling time step, but the first-order correction turned out to be sufficient in the present case.
This can be seen from the blue crosses in Fig.~\ref{fig5}(a), which are in good agreement with the theoretical predictions up to a sampling time step $\Delta t = \unit[5.0]{ms}$. The estimator for the heat flows analogous to~\eqref{eq:estimatorf} delivers equally good results, as shown in Fig.~\ref{fig5}(b).


%

\end{document}